\documentclass[reprint,showkeys,twocolumn]{revtex4-2}
\pdfoutput=1
\usepackage{graphicx}
\usepackage{bm}
\usepackage{amsmath}
\usepackage{xcolor}
\usepackage[colorlinks=true,linkcolor=blue,urlcolor=blue,citecolor=blue]{hyperref}

\usepackage{amssymb}
\usepackage{epsfig}
\usepackage{graphicx}
\usepackage[colorlinks=true,linkcolor=blue,urlcolor=blue,citecolor=blue]{hyperref}

\newcommand{\rcm}{\mbox{cm$^{-1}$}}
\newcommand{\Ss}{$^1\Sigma^{+}_{\mathrm{u}}$}
\newcommand{\Ps}{$^1\Pi_{\mathrm{u}}$}
\newcommand{\St}{$^3\Sigma^{+}_{\mathrm{u}}$}
\newcommand{\Pt}{$^3\Pi_{\mathrm{u}}$}

\begin{document}

	\title{The C\Ps\ state of potassium dimer revisited: an~extensive study by
		polarisation labelling spectroscopy method}

	\author{Wlodzimierz Jastrzebski}
	\email{jastr@ifpan.edu.pl}
	\author{Jacek Szczepkowski}	
	\affiliation{Institute of Physics, Polish Academy of Sciences,
		al.~Lotnik\'{o}w~32/46, 02-668~Warsaw, Poland}
	\author{Pawel Kowalczyk}
	\email{Pawel.Kowalczyk@fuw.edu.pl}
	\affiliation{Institute of Experimental Physics, Faculty of Physics,
		University of Warsaw, ul.~Pasteura~5, 02-093~Warszawa, Poland}%

	\begin{abstract}
		The polarisation labelling spectroscopy technique was applied to study the C\Ps\ $\leftarrow$ X$^1\Sigma^{+}_{\mathrm{g}}$ band system in potassium dimer. About 1100 new rotationally resolved molecular lines were measured in the $22100-24100$~\rcm\ spectral range. Perturbations of the lowest vibrational levels of the C state were localised and their origin discussed. A set of Dunham coefficients was deduced to fit the unperturbed levels of the C\Ps\ state with $0 \le v \le 38$ and $18 \le J \le 101$ and the potential energy curve of the state was constructed.
	\end{abstract}
	
	\date{\today}
	
	\maketitle
\section{Introduction}
\label{}

The electronic structure of potassium dimer is generally well known, both experimentally and theoretically. The ground X$^1\Sigma^{+}_{\mathrm{g}}$ state as well as several excited electronic states, particularly of singlet ungerade symmetry, have been observed and analysed in a number of works \cite{XRoss,XBHei,ALyyra,dmmy,3S3Pmy,EPmy,4PStwalley,CPmy,multimy,6S7Pmy,EKato,Ferber}. The \textit{ab initio} calculations have been performed in parallel \cite{All1,All2} providing interpretation or hints for measurements. Only a few excited states have escaped detailed experimental study, and among them, surprisingly, is the relatively low lying C(2)\Ps\ state. For the last time it was investigated nearly 30 years ago \cite{CPmy} but the resulting molecular constants were based only on about 300 observed spectral lines, the number not up to the standards of contemporary spectroscopic works. In the present experiment we extend the database related to this state, amounting now to more than 1400 transitions measured in the C\Ps\ $\leftarrow$ X$^1\Sigma^{+}_{\mathrm{g}}$ band system, identify perturbations affecting the bottom part of the C state and provide a set of Dunham coefficients reproducing positions of unperturbed energy levels in C\Ps\ up to $v'=38$ with an accuracy better than 0.1 \rcm. The origin of the observed perturbations is also discussed.

\section{Experimental}

We have used the V-type optical-optical double resonance polarisation labelling spectroscopy technique with a fixed frequency probe laser and tuneable pump one. A detailed description of the method can be found elsewhere \cite{PLS,PRA}. The experimental setup was much the same as in our recent works. Briefly, the K$_2$ molecules were produced in a stainless steel, linear heat pipe oven loaded with about 2~g of metal (natural isotope composition). Argon at pressure of 4~Torr was used as a buffer gas and the centre of the heat pipe maintained at $280^{\circ}$~C. Molecules were interacting simultaneously with nearly parallel beams of two independent lasers, both beams travelling in the same direction through the sample. The spectrally narrow, probe laser beam originating from a cw single mode ring dye laser (Coherent 899 pumped by Sprout DPSS green laser, $80-180$~mW power on DCM) was tuned to chosen transitions in the B\Ps $\leftarrow$ X$^1\Sigma^{+}_{\mathrm{g}}$ system of potassium dimer, selected on the basis of precise molecular constants of both the X and B states \cite{XBHei}, thus ``labelling'' rovibrational levels in the ground state. The ring dye laser wavelength was actively stabilised by coupling to High Finesse WS7 wavemeter with a PID controller. The pump laser, a pulsed dye laser with 0.1~\rcm\ spectral width and working on Coumarine C440 and Stilben 3 dyes (Lumonics HD-500 pumped by a Light Machinery excimer laser) was scanned across the investigated spectral range of the C\Ps $\leftarrow$ X$^1\Sigma^{+}_{\mathrm{g}}$ transitions in K$_2$. The $M$-selective optical pumping process led to a partial orientation of the molecules and the sample became optically anisotropic. Therefore in cases when both pump and probe beams interacted with the same rovibrational level in the ground electronic state, the state of polarisation of the probe beam was altered and this was detected using a pair of crossed polarisers placed on both sides of the sample. The wavenumber calibration of the scanning pump laser was obtained by simultaneous recording of the optogalvanic spectra of argon and neon as well as by sampling transmission fringes from a Fabry-P\'{e}rot etalon with FSR = 1~\rcm. In this way the absolute accuracy of 0.1~\rcm\ was achieved in positions of the observed spectral lines. All relevant signals were stored in a computer for further analysis.

\section{Analysis}

\begin{figure*}[htb!]
%	\centering \linewidth

		\includegraphics[width=0.9\textwidth]{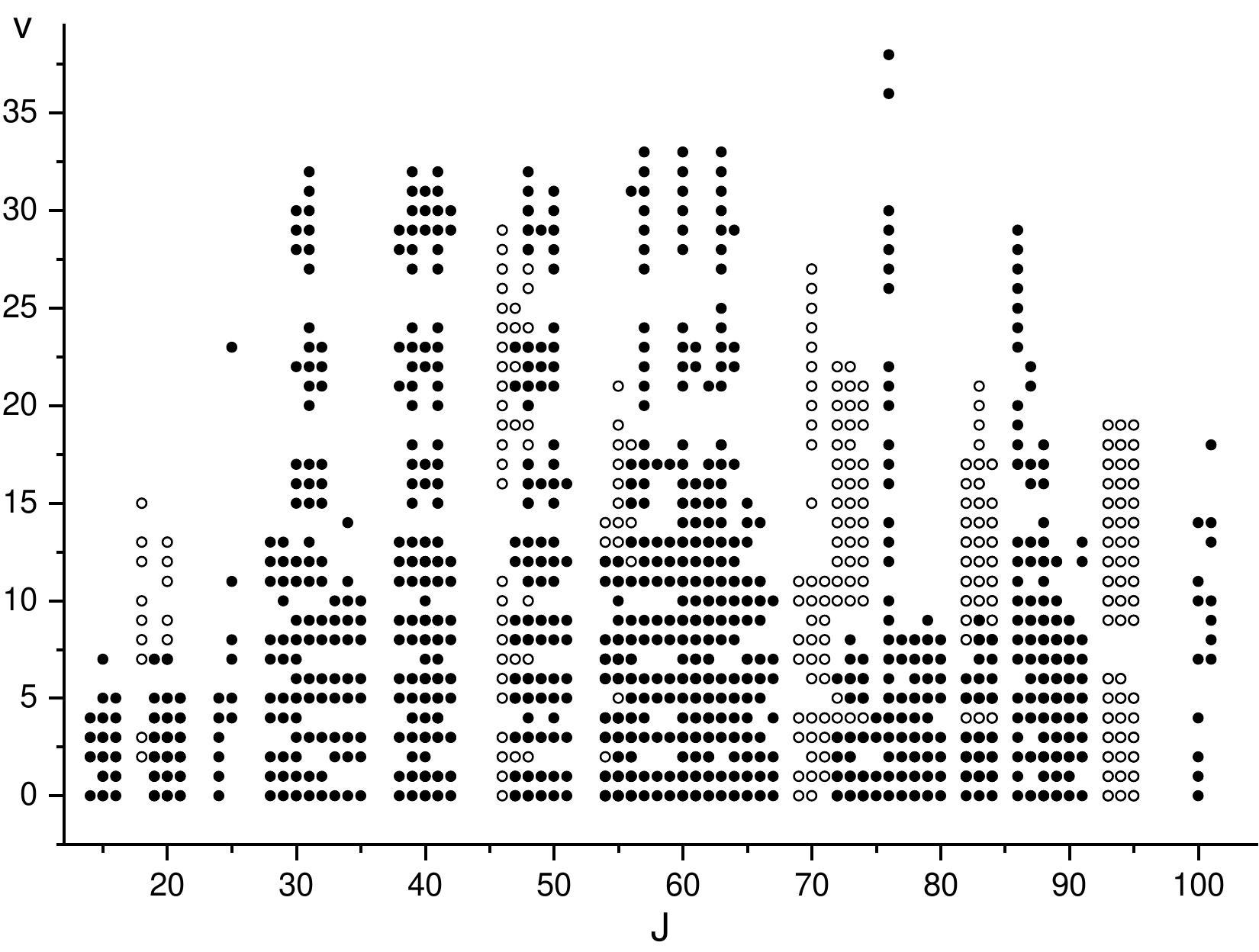}
	\caption {Range of $v$ and $J$ levels observed in the C\Ps\ state of K$_2$ in the present experiment (full circles) and in the previous one \cite{CPmy} (open circles).}
	\label{data}
\end{figure*}

The rotationally resolved excitation spectra of the C\Ps $\leftarrow$ X$^1\Sigma^{+}_{\mathrm{g}}$ band system were recorded in the range $22100-24100$~\rcm. By labelling levels $v''=0-10$, $J''=19-101$ in the ground state we were able to observe vibrational levels $v'=0-38$ in the C\Ps\ state. Altogether 1470 spectral lines were assigned and analysed (including 365 lines resulting from our previous experiment \cite{CPmy}), among them 81 lines corresponding to the rare $^{39}$K$^{41}$K isotopologue. Comparing to Ref. \cite{CPmy}, the present data field (Figure~\ref{data}) displays much better coverage of rotational levels in general and of vibrational levels for $v' > 15$. 

\begin{figure*}[!h]
%	\centering
		\includegraphics[width=0.9\linewidth]{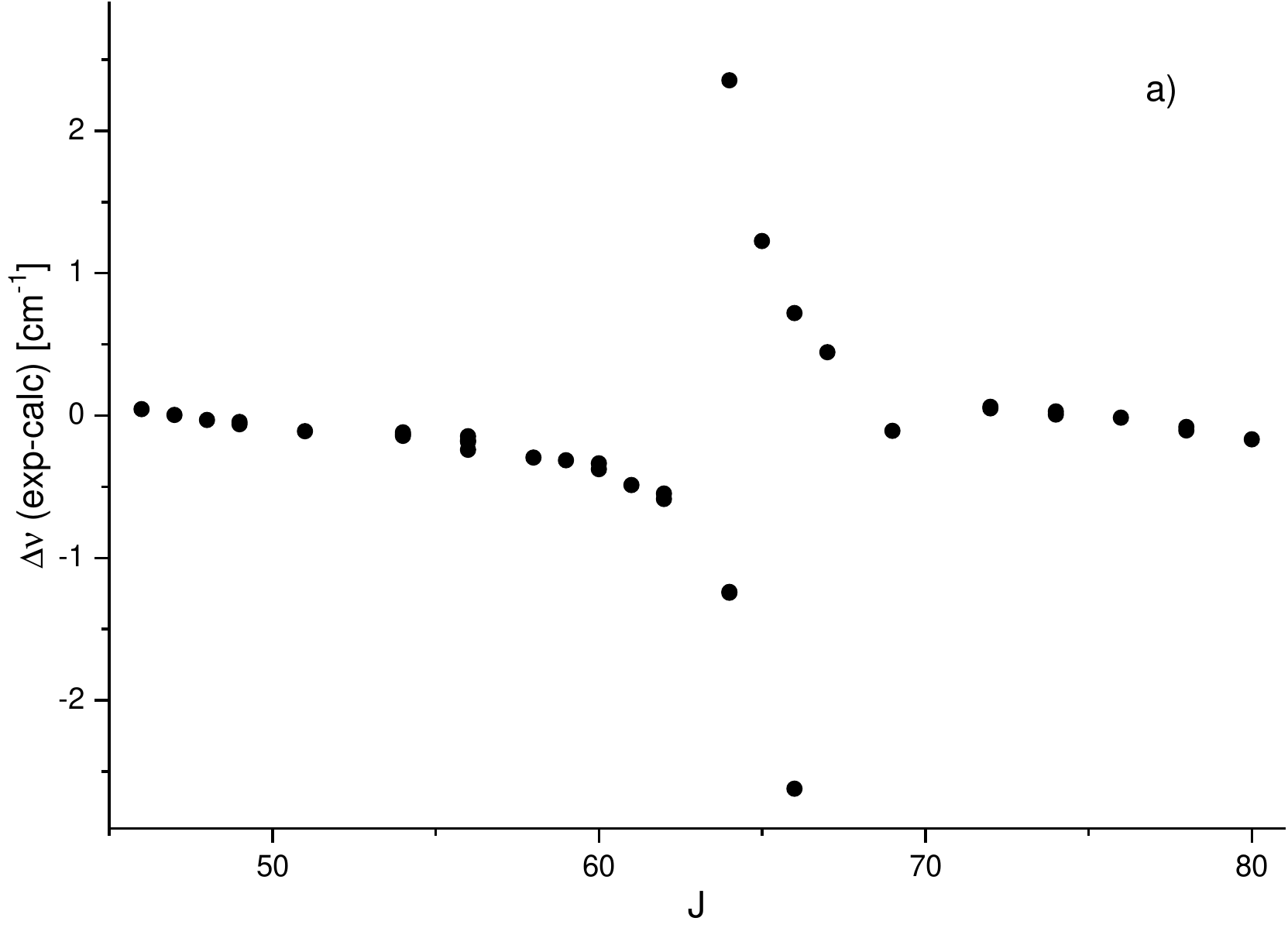}
		\includegraphics[width=0.9\linewidth]{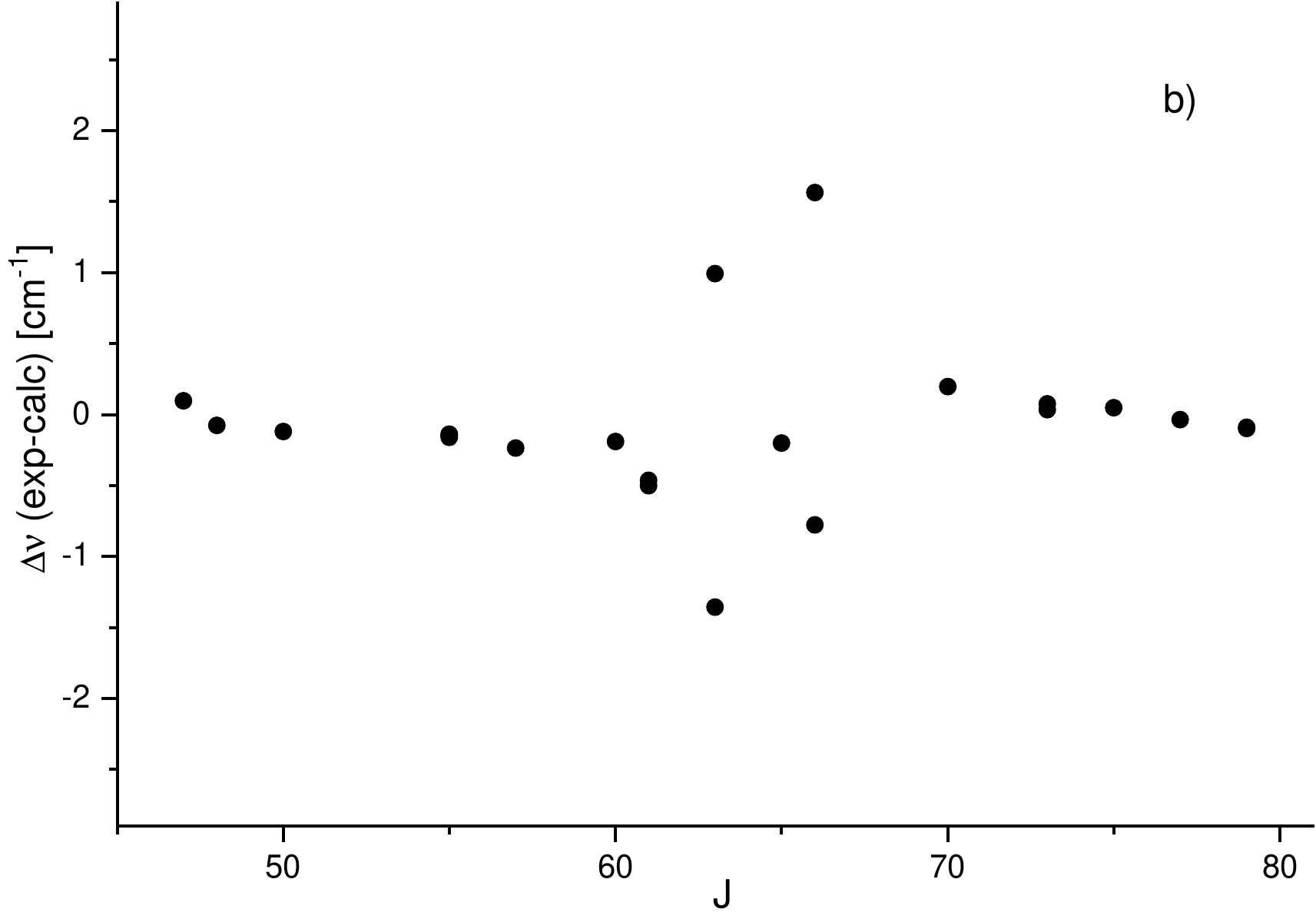}
	\caption {Differences between the measured and expected energies of rovibrational
		levels $v=0$, $45 < J < 81$ in the C\Ps\ state for (a) $e$ parity and (b) $f$ parity levels.}
	\label{ef}
\end{figure*}

\begin{figure*}[!h]
		%\centering
	\includegraphics[width=0.95\linewidth]{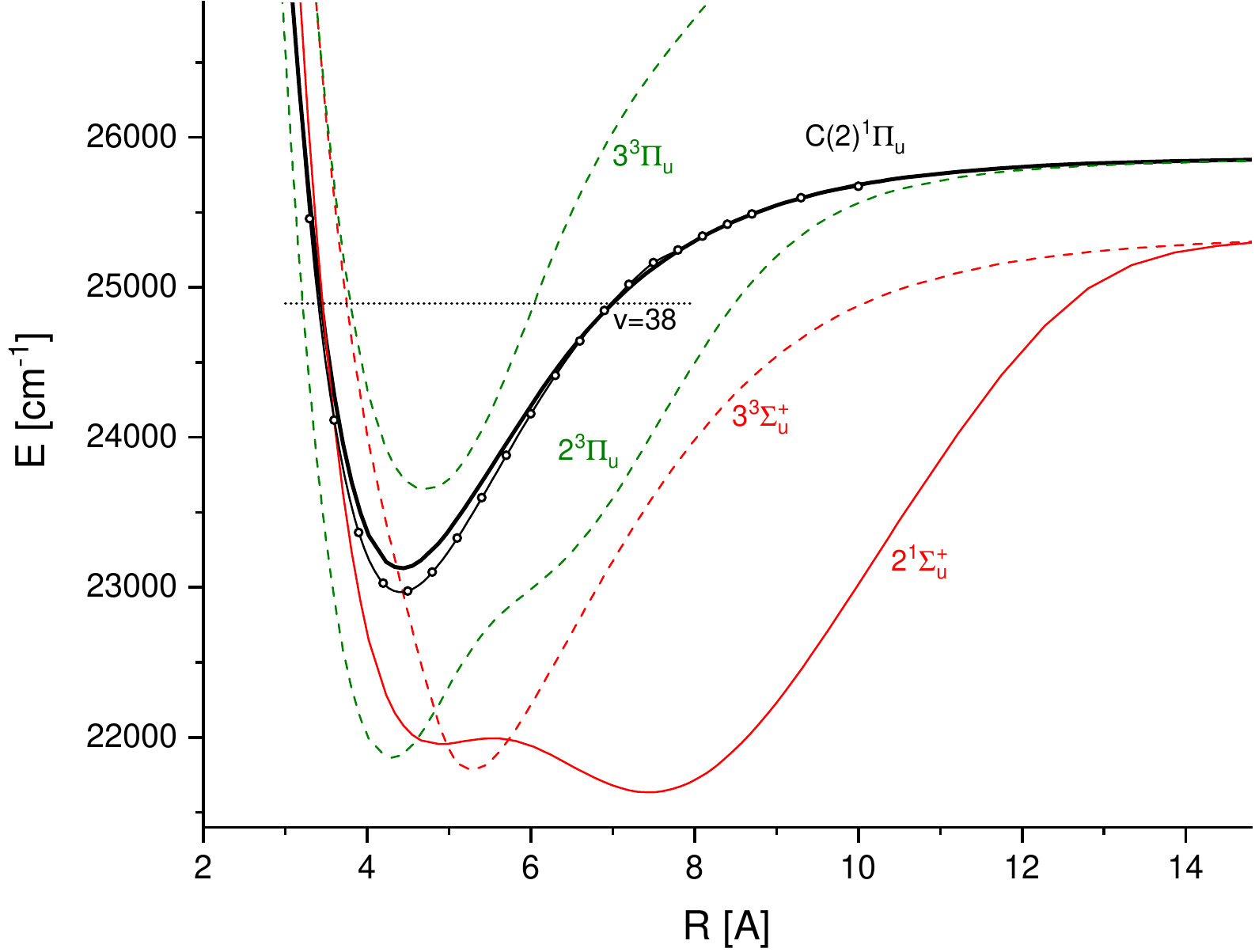}
	\caption {(Colour online) Theoretical potential curves of the electronic states of K$_2$ relevant to the present experiment~\cite{All1}, of singlet (solid lines) and triplet (dashed lines) symmetry. The experimental IPA curve (open circles) is shown alongside. The calculated points are connected with lines only to guide the eye. The dotted line denotes position of the highest observed vibrational level in the C state, $v=38$.}	\label{krzywe}
\end{figure*}

The experimental line frequencies were converted to term values of rovibronic levels in the C\Ps\ state by adding them to the ground state term values, obtained from highly accurate molecular constants of Ref. \cite{XBHei}. These data were then fitted by a least squares procedure with an equation of a form

\begin{equation}
T(v,J)=T_e+\sum_{m,n}Y_{mn}(v+0.5)^{m}[J(J+1)-1]^{n}+\delta
qJ(J+1) \mbox{ ,} \label{1}
\end{equation}

\noindent where $T_e$ denotes energy corresponding to the minimum of the potential energy curve  of the C\Ps\ state, $Y_{mn}$ are the Dunham coefficients for this state, $q$ measures the magnitude of the lambda doubling and $\delta$=0 or 1 corresponds to $f$ levels which give rise to
the Q lines or $e$ levels giving rise to P and R lines, repectively. Original fits, including all measured term values, revealed that at some energy regions considerable discrepancies occurred between the experimental energies and these calculated from the obtained molecular constants (see Figure~\ref{ef}), particularly for low vibrational levels of the C\Ps\ state. Theoretical calculations by Magnier \textit{et al.} \cite{All1} (Figure~\ref{krzywe}) predict the presence of three other electronic states in the investigated range of energies: the double minimum 2\Ss\ and two states of triplet symmetry, 2\Pt\ and 3\St. However, because of very low values of vibrational overlap integrals between C\Ps\ and the first two states mentioned above, their influence can be neglected. This leaves the 3\St\ state, a conclusion which is supported by the pattern of perturbation displayed in Figure~\ref{ef}. As rotational levels of a \St\ state are threefold degenerated (so called $F_1$, $F_2$ and $F_3$ levels, two of them corresponding to $f$ parity and one to $e$ parity), the $e$ levels of the C\Ps\ state are perturbed once but $f$ levels twice, for close values of $J$ quantum number \cite{Kovacs}. (Note that in most cases our measurements were made for rotational levels too sparse to see a difference between perturbation of $e$ and $f$ levels.)

It should be mentioned at this juncture that the C\Ps\ state was a subject of the unpublished thesis by Katern \cite{Katern}, who observed several C$\leftarrow$X bands under Doppler-free, molecular beam conditions, albeit with resolution limited by a linewidth of the employed laser to 0.04~\rcm. Although we have an access to his data, apparent problems with calibration of his spectra in terms of absolute frequencies make them unusable. Indeed, we found that positions of lines listed by Katern in some spectral regions differ systematically from our measurements by up to 0.5~\rcm\, and consequently we discarded them from the data field. Therefore we base our analysis solely upon the present measurements and those made by us in our previous experiment. We also note that Katern erroneously attributed perturbation of the C state to the 2\Pt\ state, which is ruled out by reasons stated above.

After recognising the regions of perturbations, the set of unperturbed levels proved self-consistent and allowed to determine molecular constants of the C state, listed in Table~\ref{table:Tab1}. Based on them, the Rydberg-Klein-Rees (RKR) potential energy curve was constructed. The RKR curve was further refined by a fit to a pointwise potential, using the inverted perturbation approach (IPA) method \cite{IPA}. For this purpose the experimental potential was extended below 3.4~\AA\ and above 7~\AA\ by the theoretical curve from Ref. \cite{All1} to ensure proper boundary conditions for solving the Schr\"{o}dinger equation for the investigated range of vibrational levels. The grid points defining the potential are listed in Table~\ref{table:Tab2}. To interpolate the value of the potential to an arbitrary middle point the natural cubic spline \cite{Boor} should be used, that is the second derivatives at the first and last point should be set to zero. The presented potential reproduces the term values of 1329 unperturbed levels out of a total of 1470 measured with accuracy better than 0.09~\rcm\ (the Schr\"{o}dinger equation solved in a mesh of 4000 points between 3.0 and 10.0~\AA).

\section{Discussion}

According to Refs. \cite{All1,All2} the dissociation limit of the C(2)\Ps\ state is 4s+3d$_{5/2}$. Taking into account the ground state dissociation energy $4451 \pm 1.5$~\rcm\ \cite{Amiot}, separation of the 4s and 3d$_{5/2}$ levels in atomic potassium 21534.7~\rcm\ \cite{Radzig} and $T_e$ value from Table 1, the dissociation energy of the C state can be calculated as $3018 \pm 2$~\rcm. Therefore our measurements cover about two-thirds of the potential well depth of the C\Ps\ state. The experimental IPA curve is presented alongside the theoretical ones in Figure~\ref{krzywe}. Despite a difference in $T_e$ values by 160~\rcm\ and in potential depth by 278~\rcm\ (see Table~\ref{table:Tab1}), the overall accuracy of the theoretical curve for the C state is remarkable.  

\begin{table}
\centering \caption{Molecular constants (in \rcm, except for
$R_e$ in \AA) for the C\Ps\ state of $^{39}$K$_2$ determined
in the present work. The numbers in parentheses correspond to one
standard deviation in units of the last digits, rms stands for the
root mean square deviation of the fit. The salient constants are compared with theoretical predictions of Magnier \textit{et al.} \cite{All1}}
\label{table:Tab1}\vspace*{0.5cm}
\begin{tabular}{ccc}

\hline 
 Constant & Present work & Ref. \cite{All1} \\
 \hline
     $T_e$  &  $22967.71(2)$ & 23128 \\
     $Y_{10}$ & $62.7611(6)$ & 60.52 \\
     $Y_{20}$ & $-0.36502(65)$& \\
     $Y_{30}$ & $0.18607(43) \times 10^{-2}$& \\
     $Y_{40}$ & $-0.2498(45) \times 10^{-4}$ &\\
     $Y_{01}$ & $0.0444270(48) $ & 0.0443 \\
     $Y_{11}$ & $-0.28584(50) \times 10^{-3}$& \\
     $Y_{21}$ & $-0.686(18) \times 10^{-6}$ &\\
     $Y_{02}$ & $-0.9432(48) \times 10^{-7}$ & \\
     $q$ & $0.184(11) \times 10^{-4}$ &\\
     $D_e$ & 3018(2)   & 2740 \\
     $R_e$ & 4.413(1)   & 4.42\\
     rms & 0.08 &\\
\hline \
\end{tabular}
\end{table}

\begin{table}
	
	\centering
	\caption{The rotationless IPA potential energy curve of
		the C\Ps\ state in K$_2$.}
	\label{table:Tab2}
	\begin{tabular*}{0.99\linewidth}{@{\extracolsep{\fill}}rrrr}
		&&&\\hline
		R [\AA] & U [cm$^{-1}$]& R [\AA] & U [cm$^{-1}$]\\ \hline
		&&& \\
		
3.0 & 27595.6302  &  6.3 &  24413.1223 \\
3.3 & 25457.7110  &  6.6 &  24643.2247 \\
3.6 & 24113.1393  &  6.9 &  24845.1336 \\
3.9 & 23365.4462  &  7.2 &  25020.2613 \\
4.2 & 23027.4756  &  7.5 &  25165.2485 \\
4.5 & 22975.4758  &  7.8 &  25249.2490 \\
4.8 & 23102.7288  &  8.1 &  25341.2906 \\
5.1 & 23328.0074  &  8.4 &  25419.2712 \\
5.4 & 23597.9044  &  8.7 &  25488.6416 \\
5.7 & 23880.5139  &  9.3 &  25595.8305 \\
6.0 & 24156.5145  & 10.0 &  25672.5874 \\
		
		&&&\\\hline

	\end{tabular*}
\end{table}

\begin{figure*}[htb!]
  \centering
  	\includegraphics[width=0.9\linewidth]{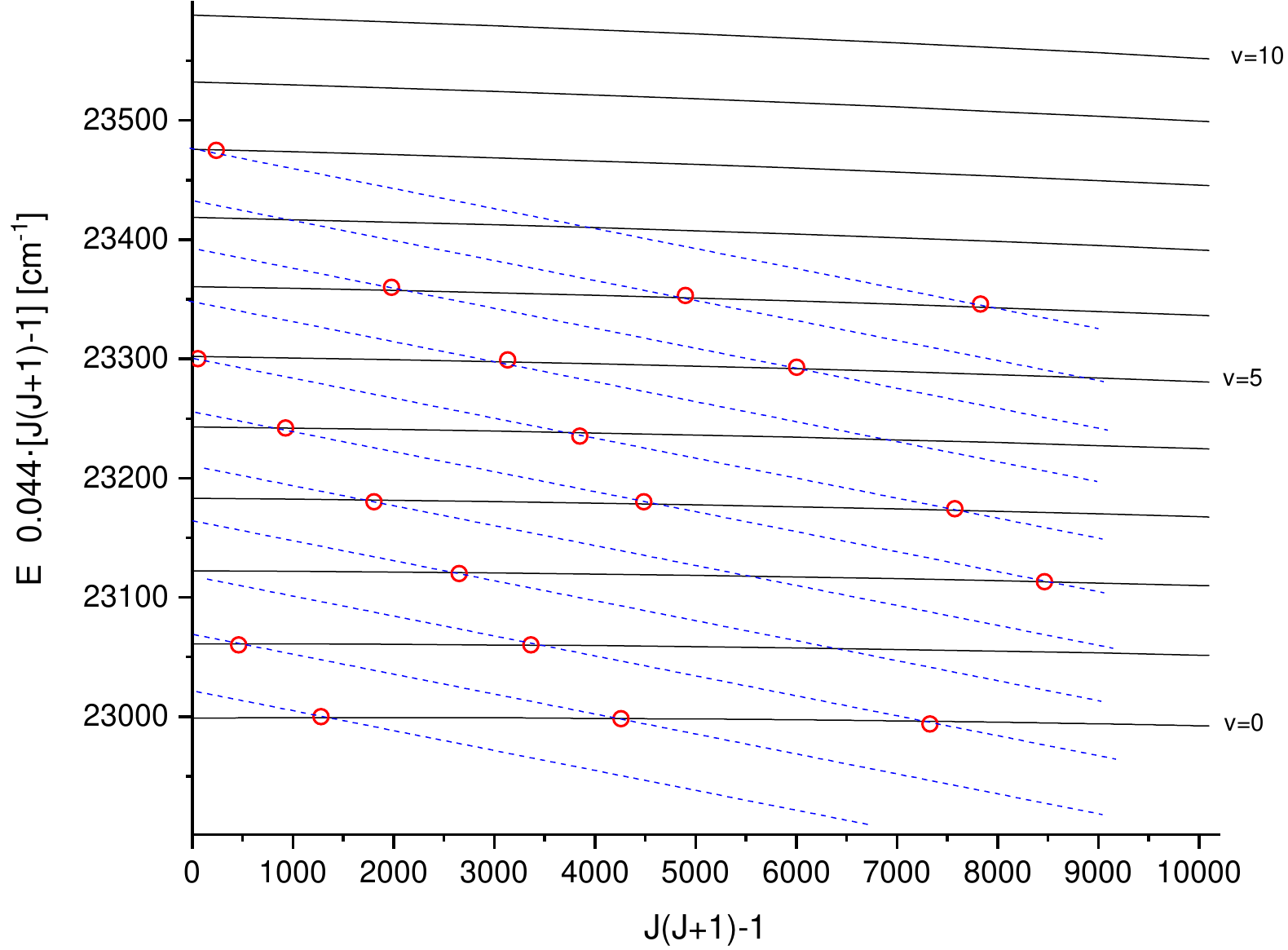}
 \caption {(Colour online) Reduced term values $E_{red}=E-0.044 \times [J(J+1)-1]$ \rcm\ of part of the observed rovibrational levels in the C\Ps\ state in K$_2$ calculated from the molecular constants listed in Table I (black solid lines) plotted against $J(J+1)-1$. Open circles (red online) indicate regions of noticeable perturbations of the C state. The dashed (blue) lines represent approximate term values of the perturbing state.}
 \label{terms}
   \end{figure*}

   In Figure~\ref{terms} we display energies of the rotational levels in the C state (in a form of the reduced term values plot) with approximate location of the observed perturbations. An important information is that the perturbations cannot be traced beyond $v'=8$. This confirms once again that the 3\St\ state is the perturber as its potential curve crosses sharply the C state potential at its very bottom, providing good vibrational overlap there, but as the energy increases, the overlap gradually gets worse (cf. Figure~\ref{krzywe}). It is noticeable that the perturbation regions can be connected by a set of (nearly) straight and (nearly) equidistant lines (blue dashed lines in Figure~\ref{terms}), apparently corresponding to subsequent vibrational levels of the perturber. From a distance of lines in Figure~\ref{terms} we can infer separation of successive vibrational levels of the perturber $\Delta G_v \approx  45$~\rcm\ and from their slope rotational constant $B_v \approx 0.0282$~\rcm. These values are again consistent with theoretical calculations \cite{All1}, predicting for the 3\St\ state in this energy range $\Delta G_v = 42.5 - 45.5$ \rcm\ and $B_v \approx 0.027$~\rcm.
   
   Suplementary data associated with this article (wavenumbers of the observed spectral lines, energies of the C state rovibrational levels, the RKR and IPA potential energy curves) can be found in its online version as well as at the address \url{http://dimer.ifpan.edu.pl}.

\section{Acknowledgements}

We gratefully acknowledge partial financial support from the National Science Centre of Poland (Grant No. 2021/43/B/ST4/03326).


\begin{thebibliography}{25}

\bibitem{XRoss}
A. J. Ross, P. Crozet, J. D’Incan, C. Effantin, J. Phys. B 19, L145--L148
(1986).

\bibitem{XBHei}
J. Heinze, U. Sch\"{u}hle, F. Engelke, C. D. Caldwell, J. Chem. Phys.
87, 45–-53 (1987).

\bibitem{ALyyra}
G. Jong, L. Li, T. J. Wang, W. C. Stwalley, J. A. Coxon, M. Li, A. M.
Lyyra, J. Mol. Spectrosc. 155, 115--135 (1992).

\bibitem{dmmy}
W. Jastrzebski, W. Jasniecki, P. Kowalczyk, R. Nadyak,
A. Pashov, Phys. Rev. A 62, 042509 (2000).

\bibitem{3S3Pmy}
W. Jastrzebski and P. Kowalczyk, Phys.Rev. A51, 1046--1051 (1995).

\bibitem{EPmy}
W. Jastrzebski and P. Kowalczyk, Chem. Phys. Lett. 227, 283--286 (1994).

\bibitem{4PStwalley}
J. T. Kim, C. C. Tsai, W. C. Stwalley, J. Mol. Spectrosc. 171, 200--209 (1995).

\bibitem{CPmy}
I. Jackowska, W. Jastrzebski, P. Kowalczyk, J.Phys. B29, L561--L564 (1996).

\bibitem{multimy}
I. Jackowska, W. Jastrzebski, P. Kowalczyk, J. Mol. Spectrosc. 185, 173--177 (1997).

\bibitem{6S7Pmy}
A. Grochola, W. Jastrzebski, P. Kowalczyk, S.Magnier, M.Aubert-Fr\'{e}con, J. Mol. Spectrosc. 224, 151--156 (2004).

\bibitem{EKato}
P. Kowalczyk, S. Kasahara, Md. H. Kabir, H.Kat\^{o}, J. Mol. Spectrosc. 220, 162--169 (2003).

\bibitem{Ferber}
I. Klincare, A. Lapins, M. Tamanis, R. Ferber, A. Zaitsevskii, E. A. Pazyuk and A. V. Stolyarov, J. Chem. Phys. 160, 064307 (2024).

\bibitem{All1}
S. Magnier, M. Aubert-Fr\'{e}con, A. R. Allouche, J. Chem. Phys. 121, 1771-–1781 (2004).

\bibitem{All2}
A. Jraij, A. R. Allouche, S. Magnier, M. Aubert-Fr\'{e}con, J. Chem. Phys. 130, 244307 (2009).

\bibitem{PLS}
R. Ferber, W. Jastrzebski, P. Kowalczyk, J. Quant. Spectrosc. Radiat. Transfer 58, 53--60 (1997).

\bibitem{PRA}
A. Pashov, P. Kowalczyk, W. Jastrzebski, Phys. Rev. A100, 012507 (2019).

\bibitem{Kovacs}
I. Kovacs, Rotational Structure in the Spectra of Diatomic Molecules, American Elsevier, New York, 1969.

\bibitem{Katern}
A. Katern, Ph.D. Thesis, Universit\"{a}t Bielefeld, 1988.

\bibitem{IPA}
A. Pashov, W. Jastrzebski, P. Kowalczyk, Comput. Phys. Commun. 128, 622--634 (2000).

\bibitem{Boor}
C. De Boor, A Practical Guide to Splines, Springer, Berlin, 1978.

\bibitem{Amiot}
C. Amiot, J. Mol. Spectrosc. 146, 370--382 (1991).

\bibitem{Radzig}
A. A. Radzig, B. M. Smirnov, Reference Data on Atoms, Molecules and Ions, Springer, Berlin, 1985.

\end{thebibliography}


\end{document}